\documentclass[reprint,
superscriptaddress,
 amsmath,amssymb,
 aps,
]{revtex4-1}

\usepackage{graphicx}
\usepackage{dcolumn}
\usepackage{systeme}
\usepackage{bm}
\usepackage{xcolor}
\usepackage{wrapfig}

\begin{document}

\preprint{APS/123-QED}

\title{Experimental realization of Lorentz boosts of space-time wave packets}

\author{Murat Yessenov}
\email{yessenov@knights.ucf.edu}
\affiliation{CREOL, The College of Optics \& Photonics, University of Central Florida, Orlando, FL 32816, USA}
\author{Miguel Romer}
\affiliation{CREOL, The College of Optics \& Photonics, University of Central Florida, Orlando, FL 32816, USA}
\author{Naoki Ichiji}
\affiliation{Institute of Industrial Science, The University of Tokyo, 4-6-1 Komaba, Meguro-Ku, Tokyo 153-8505, Japan}
\author{Ayman F. Abouraddy}%
\affiliation{CREOL, The College of Optics \& Photonics, University of Central Florida, Orlando, FL 32816, USA}%

\date{\today}

\begin{abstract}
It is now well-understood that a Lorentz boost of a spatially coherent monochromatic optical beam yields a so-called space-time wave packet (STWP): a propagation-invariant pulsed beam whose group velocity is determined by the relative velocity between the source and observer. Moreover, the Lorentz boost of an STWP is another STWP, whose group velocities are related by the relativistic law for addition of velocities typically associated with massive particles. We present an experimental procedure for testing this prediction in both the subluminal and superluminal regimes that makes use of spatio-temporal Fourier synthesis via a spatial light modulator. Our approach enables realizing the change in temporal bandwidth, the invariance of the spatial bandwidth, the concomitant change in the spatio-temporal wave-packet envelope, and the change in group velocity that all accompany a Lorentz boost of a monochromatic optical beam. The only consequence of the Lorentz boost not captured by this methodology is the Doppler shift in the optical carrier. This work may provide an avenue for further table-top demonstration of relativistic transformations of optical fields. 
\end{abstract}

\maketitle

\section{Introduction}

The discovery of a propagation-invariant pulsed-beam solution to Maxwell's equations by Brittingham in 1983 \cite{Brittingham83JAP}, known as the focus-wave mode (FWM), was quickly followed by the realization by B{\'e}langer \cite{Belanger86JOSAA} that such unusual field configurations can be obtained by applying a Lorentz boost to conventional optical fields. Since then, more propagation-invariant pulsed beams have been discovered, including X-waves \cite{Lu92IEEEa,Lu92IEEEb,Saari97PRL,Bowlan09OL,FigueroaBook14} and, more recently, space-time wave packets (STWPs) \cite{Kondakci16OE,Parker16OE,Kondakci17NP,Porras17OL,Efremidis17OL,Wong17ACSP2,PorrasPRA18,Wong20AS,Wong21OE,Diouf21OE,Yessenov22NC,Li22SR,Li22CP,Yessenov22AOP}. Underpinning the propagation invariance of any such pulsed beam (or wave packet) is a specific spatio-temporal spectral constraint that is intrinsic to Lorentz boosts \cite{Saari04PRE,Longhi04OE,Kondakci18PRL,Saari20JPC,Yessenov23PRA}. In other words, all propagation-invariant wave packets in free space -- classified within the general rubric of STWPs \cite{Yessenov19PRA} -- can be viewed as the result of applying a Lorentz boost to a pre-existing optical field. For example, so-called `baseband' STWPs \cite{Yessenov19PRA,Yessenov22AOP} follow from a Lorentz boost of a focused monochromatic beam \cite{Yessenov23PRA}; i.e., when an observer is in relative motion with respect to a source producing a monochromatic beam, this observer records a finite-bandwidth, propagation-invariant \textit{pulsed} beam -- rather than a monochromatic beam -- that takes the form of a baseband STWP \cite{Saari04PRE,Longhi04OE,Kondakci18PRL,Yessenov23PRA}.

Despite substantial theoretical studies of propagation-invariant wave packets since the initial discovery of Brittingham, only limited experimental progress was made in the optical domain \cite{Saari97PRL,Sonajalg97OL,Reivelt00JOSAA,Reivelt02PRE,Turunen10PO,FigueroaBook14}. In contrast, recent developments in spatio-temporal spectral synthesis has led to rapid progress in preparing STWPs, whether in the form of light sheets \cite{Kondakci17NP,Yessenov19Optica} or wave packets localized in all dimensions \cite{Yessenov22NC,Yessenov22OL}. These STWPs have led to a variety of new optical discoveries, including tunable group velocity \cite{Salo01JOA,Wong17ACSP2,Efremidis17OL,Kondakci19NC,Bhaduri19Optica,Yessenov20NC}, tunable group-velocity dispersion \cite{Malaguti08OL,Malaguti09PRA,Yessenov21ACSPhot,Hall21OLdispersion,Bejot22ACSP}, anomalous refraction \cite{Bhaduri20NP,Motz21OL}, extended propagation distances \cite{Bhaduri18OE,Bhaduri19OL}, self-healing \cite{Kondakci18OL}, and reduced speckle in biological samples \cite{Diouf22SA}. Moreover, progress in producing a wide variety of spatio-temporally structured optical fields has recently surged, including flying-focus wave packets \cite{SaintMarie17Optica,Froula18NP,Jolly20OE}, spatio-temporal vortices endowed with transverse orbital angular momentum \cite{Jhajj16PRX,Hancock19Optica,Chong20NP,Hancock21Optica,Gui22NP}, and toroidal pulses \cite{Papasimakis15NM,Wan22NP,Zdagkas22NP}. This recent resurgence in the study of STWPs has led to a revival of theoretical interest in the connection between propagation invariance and special relativity. These include elucidating the impact of the finite linewidth of a quasi-monochromatic beam on the propagation distance of the STWP after a Lorentz boost \cite{Yessenov23PRA}; obtaining closed-form expressions for a variety of propagation-invariant wave packets \cite{Ramsey23PRA}; and examining Lorentz transformations \cite{Saari20JPC} of X-waves \cite{Saari97PRL} and MacKinnon wave packets \cite{Mackinnon78FP}. In addition, recent work has shown further utility of examining Lorentz transformations of optical fields in the presence of orbital angular momentum \cite{Bliokh2012PRL,Bliokh12PRA,Bliokh2013JOpt,Smirnova2018PRA}. Moreover, examining Lorentz boosts of evanescent fields has shown that the helicity is not invariant, in contrast to the helicity of plane waves under a Lorentz boost \cite{Bliokh18PLA}.

In this paper, we present an experimental realization of Lorentz boosts implemented on optical fields by spatio-temporal Fourier synthesis using a spatial light modulator (SLM). We first present a theoretical formulation of the transformation of monochromatic plane waves, monochromatic paraxial beams, and STWPs under Lorentz boosts. In particular, we emphasize the change in group velocity and bandwidth of the optical fields in these scenarios. We find that the Lorentz boost of an STWP is another STWP.  Crucially, the group velocity of the STWP changes in different reference frames via the relativistic law of velocities for massive particles. We then describe our experimental approach that captures all the features introduced into a coherent optical field via a Lorentz boost; namely (1) the change in the temporal spectrum (or pulse width in time); (2) the change in the group velocity; (3) the invariance of the spatial bandwidth (constant beam spatial width); and (4) the consequent change in the spatio-temporal envelope of the wave packet. The only feature of the Lorentz boost that is \textit{not} captured by this experimental approach is the Doppler shift in the carrier frequency of the wave packet (which does not impact the spatio-temporal envelope). In our experiments, we confirm qualitatively and quantitatively the theoretical predictions regarding the impact of Lorentz boosts on paraxial optical beams and STWPs in both the subluminal and superluminal group-velocity regimes. This is the first experimental investigation of the connection between STWPs and special relativity and suggests a general methodology for demonstrations of relativistic transformations of optical beams.

\section{Space-time wave packets}\label{section:ConceptOfSTWPs}

STWPs are propagation-invariant wave packets whose unique propagation characteristics stem from their underlying spatio-temporal spectral structure \cite{Yessenov22AOP} in which each spatial frequency $k_{x}$ (the wave number along $x$) is tightly associated with a single temporal frequency $\omega$. We assume -- without loss of generality -- that the field is uniform along the second transverse spatial dimension $y$ ($k_{y}\!=\!0$). In the paraxial regime, propagation invariance requires that $k_{x}$ and $\omega$ are related through:
\begin{equation}\label{eq:SpectralCorrelation}
\frac{\Omega}{\omega_{\mathrm{o}}}\approx\frac{k_{x}^{2}}{2k_{\mathrm{o}}^{2}(1-\cot{\theta})},
\end{equation}
where $\Omega\!=\!\omega-\omega_{\mathrm{o}}$, $\omega_{\mathrm{o}}$ is a carrier frequency, $k_{\mathrm{o}}\!=\!\tfrac{\omega_{\mathrm{o}}}{c}$ is the associated wave number, $c$ is the speed of light in vacuum, and $\theta$ is an angle whose significance will become clear shortly. This equation indicates that the spatial bandwidth $\Delta k_{x}$ of an STWP is linked to its temporal bandwidth $\Delta\omega\!\propto\!(\Delta k_{x})^{2}$. This constraint between $k_{x}$ and $\omega$ enforces a linear relationship between $\omega$ and the axial wave number $k_{z}$:
\begin{equation}\label{Eq:LinearSpectrum}
\Omega=(k_{z}-k_{\mathrm{o}})c\tan\theta,
\end{equation}
such that the envelope of the field $E(x,z;t)\!=\!e^{i(k_{\mathrm{o}}z-\omega_{\mathrm{o}}t)}\psi(x,z;t)$ can be expressed in terms of an angular spectrum as follows:
\begin{equation}
\psi(x,z;t)=\int\!d\Omega\widetilde{\psi}(\Omega)e^{ik_{x}x}e^{-i\Omega(t-z/\widetilde{v})}=\psi(x,0;t-z/\widetilde{v}),
\end{equation}
where $\widetilde{\psi}(\Omega)$ is the Fourier transform of $\psi(0,0;t)$. The envelope travels rigidly in free space without diffraction or dispersion at a group velocity $\widetilde{v}\!=\!c\tan\theta$.

\begin{figure}[t!]
    \centering
    \includegraphics[width=8.6cm]{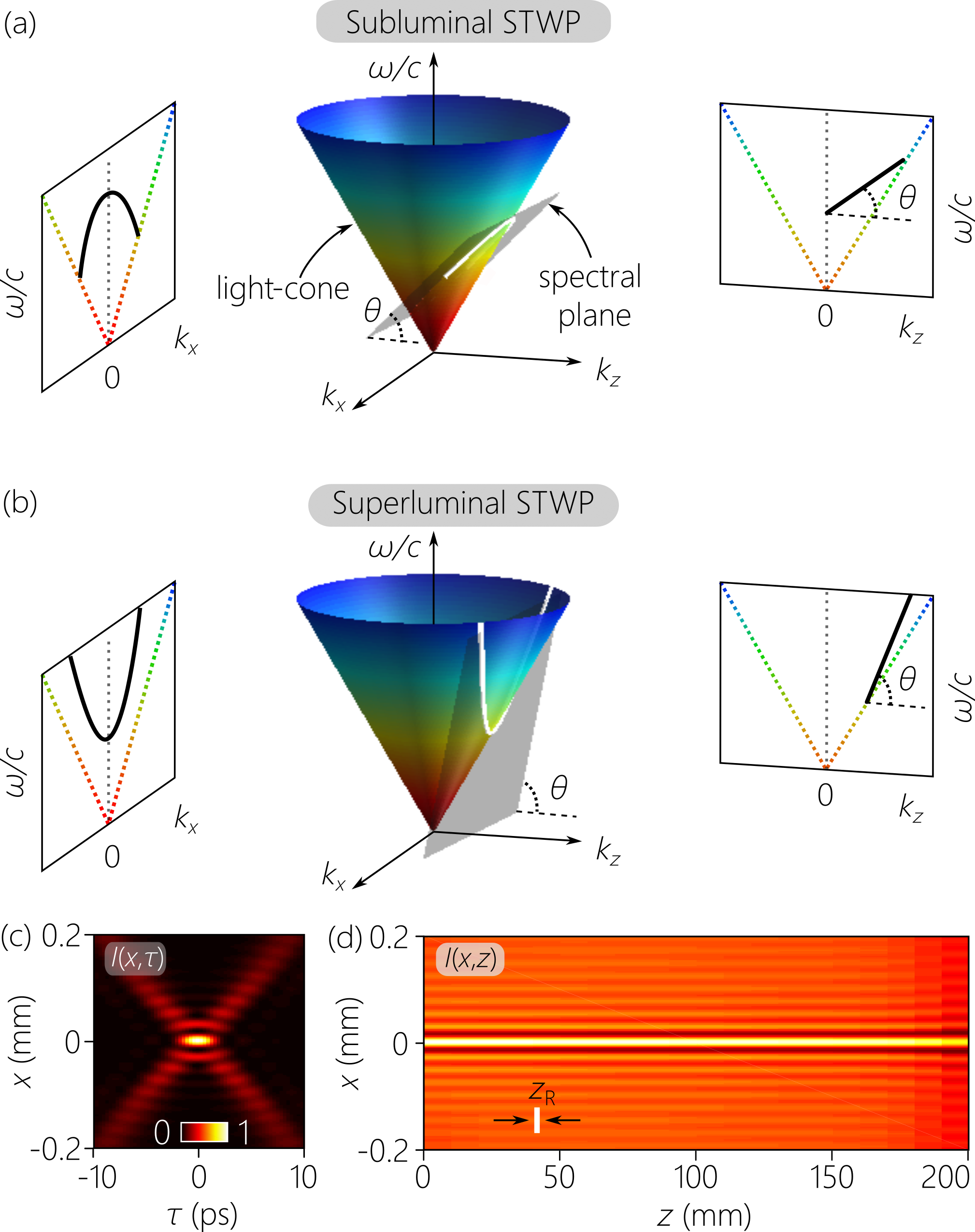}
    \caption{(a) Representation of the spatio-temporal spectrum of a propagation-invariant STWP at the intersection of the light-cone with a tilted spectral plane. The spectral tilt angle $\theta\!<\!45^{\circ}$ corresponds to a subluminal STWP. We also plot the spectral projections onto the $(k_{x},\tfrac{\omega}{c})$-plane (left; portion of an ellipse), and onto the $(k_{z},\tfrac{\omega}{c})$-plane (right; a straight line). (b) Same as (a) for a superluminal STWP. Here $\theta\!>\!45^{\circ}$, the spectral projection onto the $(k_{x},\tfrac{\omega}{c})$-plane is a hyperbola. The dotted lines in (a,b) are the light-lines $k_{z}\!=\!\pm\tfrac{\omega}{c}$. (c) Spatio-temporal intensity $I(x,0;t)$ of the STWP at a fixed axial plane $z\!=\!0$. (d) Time-averaged intensity $I(x,z)$ for an STWP. In (c,d), we have $\theta\!=\!50^{\circ}$, $\Delta\lambda\!=\!1$~nm, $\lambda_{o}\!=\!800$~nm, $\Delta k_{x}\!=\!0.16$~rad/$\mu$m, and $\Delta x\!=\!20$~$\mu$m. The white rectangle represents the Rayleigh length of a Gaussian beam with the same spatial width $\Delta x$ of the STWP.}
    \label{fig:STWPconcept}
\end{figure}

The linear relationship in Eq.~\ref{Eq:LinearSpectrum} corresponds geometrically to a plane in $(k_{x},k_{z},\tfrac{\omega}{c})$-space that is parallel to the $k_{x}$-axis and makes an angle $\theta$ (which we call the spectral tilt angle) with the $k_{z}$-axis. The spatio-temporal spectrum of an STWP lies on the conic section at the intersection of the free-space light cone $k_{x}^{2}+k_{z}^{2}\!=\!(\tfrac{\omega}{c})^{2}$ with this spectral plane, and passes through the point $(k_{x},k_{z},\tfrac{\omega}{c})\!=\!(0,k_{\mathrm{o}},k_{\mathrm{o}})$. When $0^{\circ}\!<\!\theta\!<\!45^{\circ}$ or $135^{\circ}\!<\!\theta\!<\!180^{\circ}$, the conic section is an ellipse and the group velocity $|\widetilde{v}|\!<\!c$ is subluminal. When $45^{\circ}\!<\!\theta\!<\!135^{\circ}$, the conic section is a hyperbola, and the group velocity $|\widetilde{v}|\!>\!c$ is superluminal. When $\theta\!=\!45^{\circ}$ and $\widetilde{v}\!=\!c$, the spatio-temporal spectrum is confined to the light line $k_{z}\!=\!\tfrac{\omega}{c}$ ($k_{x}\!=\!0$), corresponding to a conventional luminal plane-wave pulse. When $\theta\!=\!135^{\circ}$, the conic section is a parabola with $\widetilde{v}\!=\!-c$. Finally, when $\theta\!=\!0^{\circ}$, the iso-frequency ($\omega\!=\!\omega_{\mathrm{o}}$) conic section is a circle, corresponding to a monochromatic beam ($\widetilde{v}\!\rightarrow\!0$).

In Fig.~\ref{fig:STWPconcept}(a) we depict the locus of the spatio-temporal spectrum for a subluminal STWP $\theta\!<\!45^{\circ}$, which is a portion of an ellipse (in the vicinity of $k_{x}\!=\!0$) at the intersection of the light-cone and a tilted spectral plane. Because the spectral plane is parallel to the $k_{x}$-axis, the projection of the spatio-temporal spectrum onto the $(k_{z},\tfrac{\omega}{c})$-plane is a straight line, while that onto the $(k_{x},\tfrac{\omega}{c})$-plane is an ellipse. The slope of the projected line onto the $(k_{z},\tfrac{\omega}{c})$-plane signifies the group velocity of the STWP, and the absence of curvature implies no group-velocity dispersion of any order \cite{Kondakci17NP,Kondakci19NC}. The corresponding configuration for a superluminal STWP ($\theta\!>\!45^{\circ}$) is depicted in Fig.~\ref{fig:STWPconcept}(b). Finally, we plot an example of the spatio-temporal intensity profile $I(x,0;t)\!=\!|E(x,0;t)|^{2}$ at a fixed axial plane $z\!=\!0$ [Fig.~\ref{fig:STWPconcept}(c)]. In contrast to a conventional pulsed beam which is typically separable with respect to the spatial and temporal degrees of freedom, the STWP intensity profile is non-separable and has a characteristic X-shaped profile \cite{Yessenov22AOP}. Because such an STWP travels invariantly, its time-averaged intensity $I(x,z)\!=\!\int\!dtI(x,z;t)$ is diffraction-free, $I(x,z)\!=\!I(x,0)$ [Fig.~\ref{fig:STWPconcept}(d)]. These unique characteristics of STWPs have now been verified experimentally in detail \cite{Yessenov22AOP}.  

\section{Lorentz transformation of monochromatic plane waves} \label{section:transfSeparableLight}

\begin{figure}[t!]
    \centering
    \includegraphics[width=8.6cm]{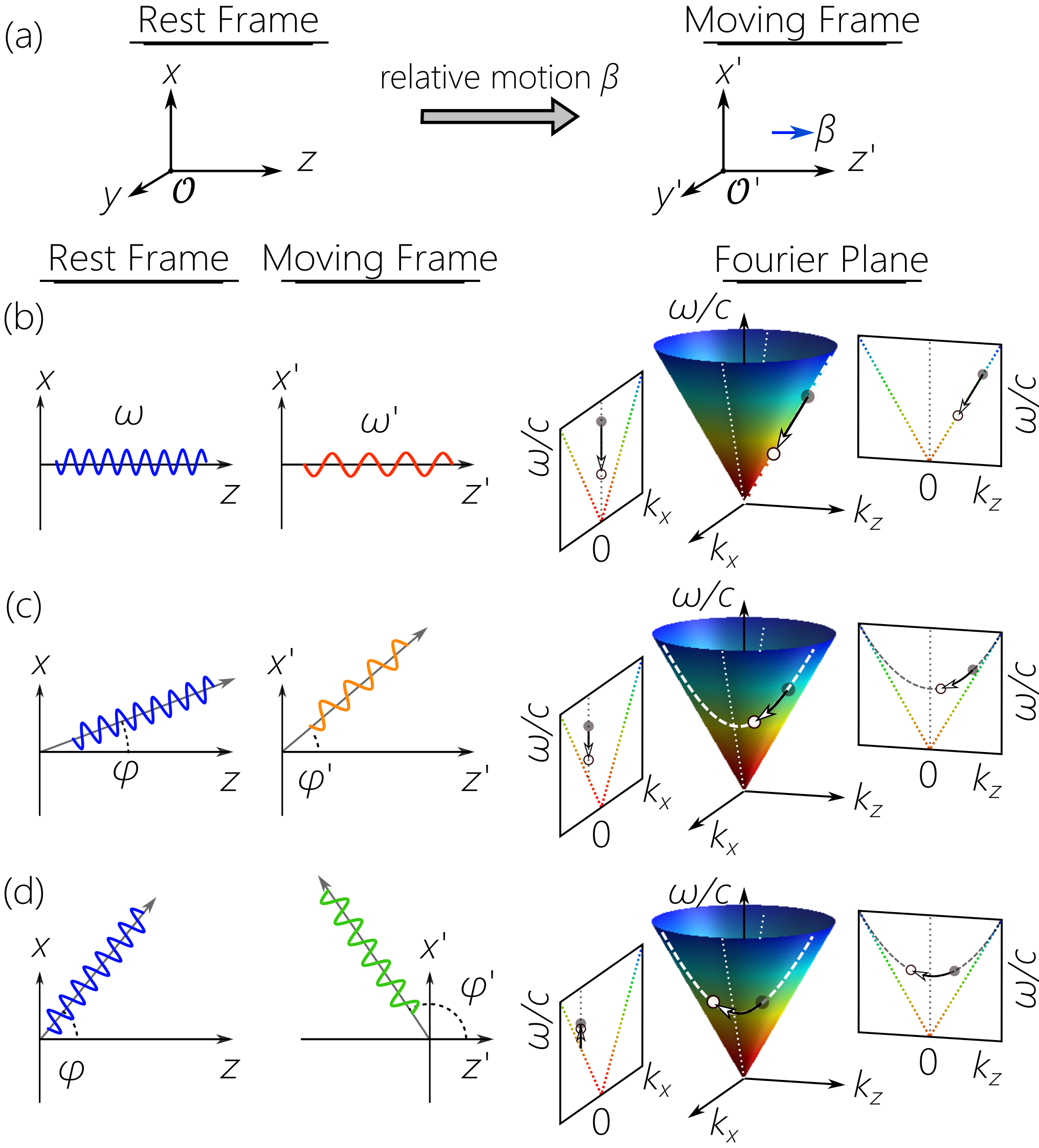}
    \caption{(a) We consider the field observed in the reference frame $\mathcal{O}'(x',z';t')$, referred to as the moving frame, when emitted from a source in the frame $\mathcal{O}(x,z;t)$, referred to as the rest frame. The frame $\mathcal{O}'$ moves at a velocity $v\!=\!\beta c$ relative to $\mathcal{O}$ along the common $z$-axis. (b) Depiction of the Lorentz boost of an on-axis MPW, along with the change in its geometrical representation as a point on the light-line $k_{z}\!=\!\tfrac{\omega}{c}$ on the spectral light-cone surface. The frequency $\omega_{\mathrm{o}}$ in $\mathcal{O}$ is Doppler-shifted to $\omega_{\mathrm{o}}'$ in $\mathcal{O}'$, and the point is displaced along the light-line. (c) Same as (b) for an off-axis MPW. The frequency is Doppler-shifted \textit{and} the propagation angle with the $z$-axis changes in $\mathcal{O}'$. Here, the point representing the MPW (whether in $\mathcal{O}$ or $\mathcal{O}'$) does not lie on the light-line. The angle-dependent Doppler shift associated with the Lorentz boost translates this point along an iso-$k_{x}$ hyperbola. (d) Same as (c) when the Lorentz boost results in a change in the sign of axial wave number $k_{z}$.}
    \label{fig:MonochrPlaneWaveTransformation}
\end{figure}

To establish the connection between STWPs and Lorentz boosts of conventional optical beams, we first summarize the changes undergone by a monochromatic plane wave (MPW) when recorded by a moving observer. The optical source emitting the MPW is assumed to be in a reference frame $\mathcal{O}(x,z;t)$, referred to hereon as the rest frame, and the observer is in the reference frame $\mathcal{O}'(x',z';t')$ that moves at a velocity $v\!=\!\beta c$ along the $+z$ direction with respect to $\mathcal{O}$, referred to hereon as the moving frame [Fig.~\ref{fig:MonochrPlaneWaveTransformation}(a)]. The Lorentz transformation relating the spectral parameters in the two frames is:
\begin{eqnarray}\label{Eq:LorentzTransFourierDomain}
k_{x}'&=&k_{x},\nonumber\\
k_{z}'&=&\gamma(k_{z}-\beta\omega/c),\nonumber\\
\omega'/c\!&=&\gamma(\omega/c-\beta k_{z}),
\end{eqnarray}
where $\gamma=1/\sqrt{1-\beta^{2}}$. An MPW of the form $e^{i(k_{x}x+k_{z}z-\omega t)}$ is represented geometrically by a point on the surface of the spectral light-cone $k_{x}^{2}+k_{z}^{2}\!=\!(\tfrac{\omega}{c})^{2}$ (not to be confused with the conventional space-time light-cone). The structure of the spectral light-cone in Fig.~\ref{fig:STWPconcept} is preserved under a Lorentz-boost: $k_{x}^{2}+k_{z}^{2}\!=\!(\tfrac{\omega}{c})^{2}\Longrightarrow k_{x}'^{2}+k_{z}'^{2}\!=\!(\tfrac{\omega'}{c})^{2}$. We therefore can represent the MPWs in both frames $\mathcal{O}$ and $\mathcal{O}'$ on the same light-cone surface.

\begin{figure*}[t!]
    \centering
    \includegraphics[width=16cm]{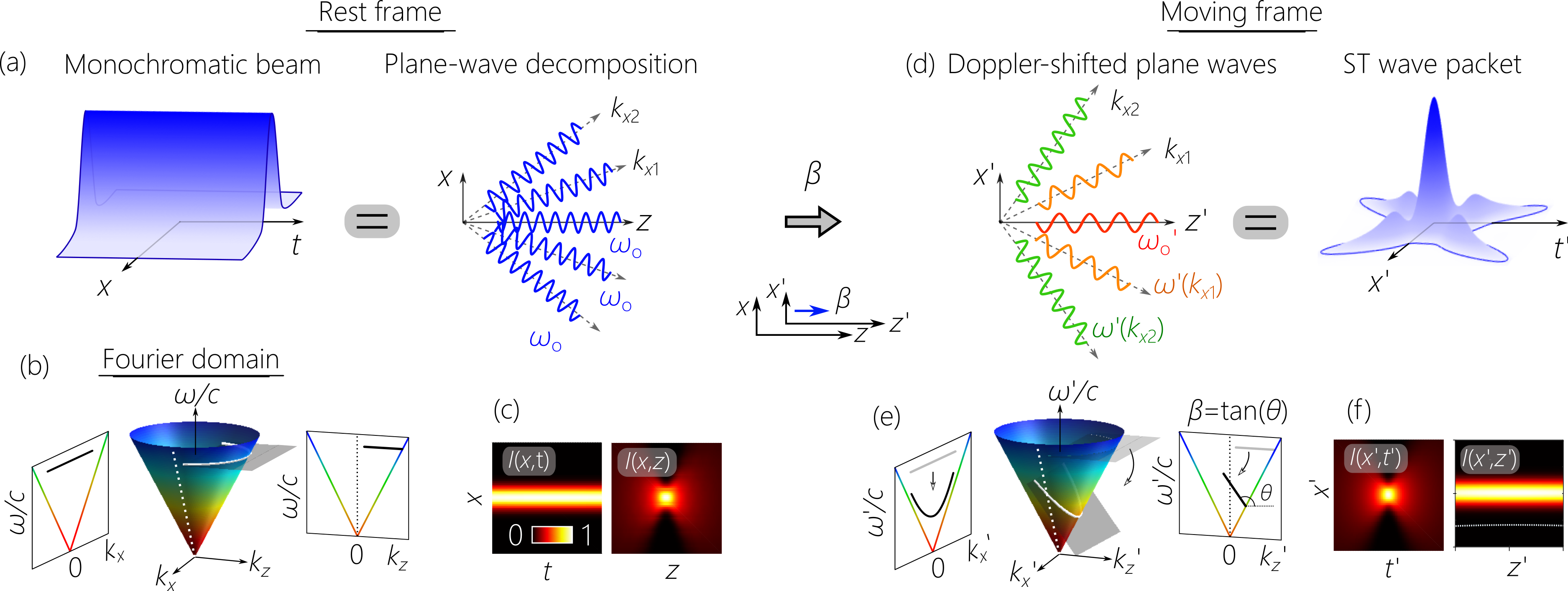}
    \caption{(a-c) A monochromatic beam at frequency $\omega_{\mathrm{o}}$ in the rest frame $\mathcal{O}$. (a) The monochromatic beam is a superposition of MPWs, all of which have the same temporal frequency $\omega_{\mathrm{o}}$, but travel at different angles with the $z$-axis. (b) The spatio-temporal spectrum of the monochromatic beam in (a) lies on the circle at the intersection of the spectral light-cone with a horizontal iso-frequency plane $\omega\!=\!\omega_{\mathrm{o}}$. The spectral projections onto the $(k_{x},\tfrac{\omega}{c})$ and $(k_{z},\tfrac{\omega}{c})$ planes are both horizontal straight lines. (c) The monochromatic beam is stationary in time $I(x,z;t)\!=\!I(x,z;0)$ at any fixed axial plane (here $z\!=\!0$), but is spatially localized at all instants of time, so that the time-averaged intensity $I(x,z)$ remains spatially localized along $z$. (d-f) STWP in the moving frame $\mathcal{O}'$ resulting from a Lorentz boost to the monochromatic beam in the rest frame $\mathcal{O}$ shown in (a-c). (d) Each plane wave in (a) undergoes a different angle-dependent Doppler shift in $\mathcal{O}'$ after the Lorentz boost, so that the beam is no longer monochromatic, and is instead \textit{pulsed} with a finite temporal bandwidth $\Delta\omega'$. (e) The spatio-temporal locus of the resulting Lorentz-boosted field lies at the intersection of the light-cone with a \textit{tilted} spectral plane, which makes an angle $\theta$ with the $k_{z'}$-axis, where $\tan{\theta}\!=\!-\beta$. The spectral projection onto the $(k_{z}',\tfrac{\omega'}{c})$-plane is a straight line making an angle $\theta$ with the $k_{z}'$-axis. The spectral projection onto the $(k_{x}',\tfrac{\omega'}{c})$-plane is a portion of an ellipse in the vicinity of $k_{x}'\!=\!0$. (f) The STWP is no longer stationary in time, but the time-averaged intensity profile is diffraction-free $I'(x',z')\!=\!I'(x',0)$.}
    \label{fig:MonochrBeamTransformation}
\end{figure*}

We consider three scenarios for an MPW at a frequency $\omega_{\mathrm{o}}$ [Fig.~\ref{fig:MonochrPlaneWaveTransformation}(b-d)]. First, if the MPW in $\mathcal{O}$ propagates along the common $z$-axis ($k_{x}\!=\!0$, represented by a point on the light-line $k_{z}\!=\!\tfrac{\omega}{c}$), it also propagates along the $z'$-axis ($k_{x}'\!=\!0)$ in $\mathcal{O}'$. The Doppler shift associated with the Lorentz boost translates the point representing the MPW along the light-line $k_{z}'\!=\!\tfrac{\omega'}{c}$ to $\omega_{\mathrm{o}}'\!=\!\omega_{\mathrm{o}}\sqrt{\tfrac{1-\beta}{1+\beta}}$ [Fig.~\ref{fig:MonochrPlaneWaveTransformation}(b)]. In the second scenario [Fig.~\ref{fig:MonochrPlaneWaveTransformation}(c)], we consider an off-axis MPW in $\mathcal{O}$ traveling at an angle $\varphi$ with the $z$-axis ($k_{x}\!=\!\tfrac{\omega}{c}\sin\varphi$). In $\mathcal{O'}$ the MPW undergoes an angle-dependent Doppler shift to $\omega_{\mathrm{o}}'\!=\!\omega_{\mathrm{o}}\gamma(1-\beta\cos\varphi)$, and the propagation angle with the $z'$-axis changes from $\varphi$ to $\varphi'$, where $\cos\varphi'\!=\!\tfrac{\cos\varphi-\beta}{1-\beta\cos\varphi}$. The off-axis MPW is represented by a point on the line-cone that does \textit{not} lie on the light-line. The invariance of the transverse wave number $k_{x}$ ensures that the Lorentz boost translates this point along an iso-$k_{x}$ hyperbola [Fig.~\ref{fig:MonochrPlaneWaveTransformation}(c)]. A different scenario emerges for an off-axis MPW when $\beta\!>\!\cos\varphi$, where the sign of $k_{z}'$ is opposite of that of $k_{z}$, $\cos\varphi'\!<\!0$, and the MPW in $\mathcal{O}'$ travels backwards along the negative $z'$-axis, whereas the MPW in $\mathcal{O}$ travels along the positive $z$-axis. In this scenario, the point representing the MPW in $\mathcal{O}$ travels along the iso-$k_{x}$ hyperbola and crosses the origin point ($k_{z}\!=\!0$) towards $k_{z}\!<\!0$ [Fig.~\ref{fig:MonochrPlaneWaveTransformation}(d)]. Because this occurs at relativistic speeds $\beta\!\rightarrow\!1$ or in the non-paraxial regime, we do not consider this special case further.

\section{Lorentz boost of monochromatic paraxial beams}\label{section:MonochrBeam}

Now consider the impact of a Lorentz boost on a \textit{monochromatic paraxial beam} produced from a source in the rest frame $\mathcal{O}$. The field $E(x,z;t)$ can be factored into a slowly varying \textit{spatial} envelope $\psi(x,z)$ and a carrier, $E(x,z;t)\!=\!e^{i(k_{\mathrm{o}}z-\omega_{\mathrm{o}}t)}\psi(x,z)$, and the envelope can be expressed as an angular spectrum:
\begin{equation}
\psi(x,z)=\int\!dk_{x}\widetilde{\psi}(k_{x})e^{ik_{x}x}e^{i(k_{z}-k_{\mathrm{o}})z},
\end{equation}
where $\widetilde{\psi}(k_{x})$ is the Fourier transform of $\psi(x,0)$. In other words, the monochromatic beam is a superposition of iso-frequency MPWs, each traveling at a different angle $\varphi$ with the $z$-axis [Fig.~\ref{fig:MonochrBeamTransformation}(a)], where $k_{x}\!=\!k_{\mathrm{o}}\sin\varphi$. Such a beam is represented by a portion of the circle (in the vicinity of $k_{x}\!=\!0$) at the intersection of the light-cone with the horizontal iso-frequency plane $\omega\!=\!\omega_{\mathrm{o}}$ [Fig.~\ref{fig:MonochrBeamTransformation}(b)]. The spectral projections onto the $(k_{x},\tfrac{\omega}{c})$ and $(k_{z},\tfrac{\omega}{c})$ planes are therefore both horizontal lines. The field is stationary in time $E(x,z;t)\!=\!e^{-i\omega_{\mathrm{o}}t}E(x,z;0)$, so that the intensity $I(x,z;t)\!=\!|E(x,z;t)|^{2}\!=\!I(x,z;0)$ is independent of $t$, whereas the intensity profile is spatially localized along the $z$-axis at any time $t$ [Fig.~\ref{fig:MonochrBeamTransformation}(c)].

In the moving frame $\mathcal{O}'$, each MPW in the monochromatic beam undergoes a \textit{different} angle-dependent Doppler shift [Fig.~\ref{fig:MonochrBeamTransformation}(d)], as described above. Therefore, the initially monochromatic beam at $\omega\!=\!\omega_{\mathrm{o}}$ acquires a \textit{finite} temporal bandwidth $\Delta\omega'$ centered at the frequency $\omega_{\mathrm{o}}'\!=\!\omega_{\mathrm{o}}\gamma(1-\beta)$ associated with the on-axis MPW ($k_{x}\!=\!0$). The mutual \textit{spatial} coherence of the initially iso-frequency MPW constituting the monochromatic beam ensures that the mutual \textit{temporal} coherence of the different-frequency MPWs after the Lorentz-boost.

Therefore, the observer in $\mathcal{O}'$ records a \textit{pulsed} beam $E'(x',z';t')\!=\!e^{i(k_{\mathrm{o}}'z'-\omega_{\mathrm{o}}'t')}\psi'(x',z';t')$ rather than a monochromatic beam, where the space-time coordinates are related through the usual formulas for the Lorentz boost: $x'\!=\!x$, $z'\!=\!\gamma(z-\beta ct)$, and $ct'\!=\!\gamma(ct-\beta z)$. In general, the spatio-temporal field of this pulsed beam in $\mathcal{O}'$ is related to that in $\mathcal{O}$ via:
\begin{equation}\label{Eq:GeneralFieldTransformation}
E'(x',z';t')=E[x',\gamma(z'+\beta ct');\gamma(t'+\beta z'/c)].
\end{equation}
For the case of a monochromatic beam in $\mathcal{O}$, the field in $\mathcal{O}'$ is $E'(x',z';t')\!=\!e^{i(k_{\mathrm{o}}'z'-\omega_{\mathrm{o}}'t)}\psi'(x',z';t')$, which is a product of a carrier and a \textit{spatio-temporal} envelope that is related to the purely \textit{spatial} envelope $\psi(x,z)$ in $\mathcal{O}$ via:
\begin{equation}
\psi'(x',z';t')=\psi[x',\gamma(z'+\beta ct')];
\end{equation}
which results in propagation invariance in $\mathcal{O}'$:
\begin{equation}
\psi'(x',z';t')=\psi'(x',0;t'-z'/\widetilde{v}),
\end{equation}
where $\widetilde{v}\!=\!-v$.

The field in $\mathcal{O}'$ has several intriguing characteristics:
\begin{enumerate}
    \item It travels rigidly in space; i.e., it is propagation invariant without experiencing diffraction or dispersion: $I'(x',z';t')=I'(x',0,t'-z'/\widetilde{v})$.
    \item It travels in $\mathcal{O}'$ at a group velocity $\widetilde{v}\!=\!-v$.
    \item Because $|\beta|\!<\!1$, the resulting wave packet in $\mathcal{O}'$ is \textit{subluminal}, $|\widetilde{v}|\!<\!c$.
    \item Its spatio-temporal profile $I'(x',0;t')$ at a fixed axial location $z'\!=\!0$ maps to the purely spatial profile of the monochromatic beam $I(x,z)$ in $\mathcal{O}$ \cite{Longhi04OE,Kondakci18PRL}. 
\end{enumerate}

Because the spatial frequencies $k_{x}$ are invariant under the Lorentz boost $k_{x}'\!=\!k_{x}$, the spatial bandwidth of the pulsed beam $\Delta k_{x}'$ in $\mathcal{O}'$ is the same as the spatial bandwidth $\Delta k_{x}$ of the monochromatic beam in $\mathcal{O}$, $\Delta k_{x}'\!=\!\Delta k_{x}$. Therefore, the \textit{spatial} beam profile at $t'\!=\!0$ at a fixed axial plane $z'\!=\!0$ in $\mathcal{O}'$, $I'(x',0;0)$ is identical to that of the monochromatic beam in $\mathcal{O}$, $I(x,0)$. In particular, the spatial beam width in this plane remains the same $\Delta x\!=\Delta x'\!\sim\!\tfrac{1}{\Delta k_{x}}$. On the other hand, the temporal bandwidth $\Delta\omega'$ in $\mathcal{O}'$ can be shown to be given by:
\begin{equation}
\Delta\omega'=\frac{\omega_{\mathrm{o}}\gamma|\beta|}{2}\left(\frac{\Delta k_{x}}{k_{\mathrm{o}}}\right)^{2},
\end{equation}
from which we can estimate the pulse width $\Delta t'\!\sim\!\tfrac{1}{\Delta\omega'}$ of the \textit{temporal} envelope $I'(0,0;t')$ at the beam center $x'\!=\!0$ and a fixed axial plane $z'\!=\!0$.

Here the change in the geometrical representation of the spatio-temporal spectrum on the light-cone surface is particularly instructive. In $\mathcal{O}$, the monochromatic beam corresponds to a portion of a horizontal, iso-frequency circle [Fig.~\ref{fig:MonochrPlaneWaveTransformation}(b)]. Each point on this circle moves under the same Lorentz boost along \textit{different} iso-$k_{x}$ hyperbolas [Fig.~\ref{fig:MonochrPlaneWaveTransformation}(c)]. The linearity of the algebraic form of the Lorentz boost guarantees that the spectral locus remains the intersection of the light-cone with a plane, which is now tilted by an angle $\theta$ (the spectral tilt angle) with the $k_{z}$-axis, where $\tan\theta\!=\!-\beta$:
\begin{equation}
\omega-\omega_{\mathrm{o}}=0\Longrightarrow\omega'-\omega_{\mathrm{o}}'=(k_{z}'-k_{\mathrm{o}}')c\tan{\theta'}
\end{equation}

The spatio-temporal spectrum now lies on a conic section (ellipse, hyperbola, or parabola), depending on the value of $\theta$. However, when $|\beta|\!<\!1$, the conic section in this scenario is always an ellipse because $0^{\circ}\!<\!\theta\!<\!45^{\circ}$ or $135^{\circ}\!<\!\theta\!<\!180^{\circ}$ ($|\widetilde{v}|\!<\!c$). Whereas the spectral projection onto the $(k_{z},\tfrac{\omega}{c})$-plane is still a straight line, albeit one that makes an angle $\theta$ with the $k_{z}$-axis, the spectral projection onto the $(k_{x},\tfrac{\omega}{c})$-plane is no longer a straight line, but is instead a portion of a conic section [Fig.~\ref{fig:MonochrBeamTransformation}(e)]. All the characteristics of the pulsed beam in $\mathcal{O}'$ are those of a baseband STWP. In other words, the conventional continuous-wave monochromatic beam in $\mathcal{O}$ is recorded as a propagation-invariant STWP in $\mathcal{O}'$. The spatio-temporal intensity profile is localized in both space and time at any fixed axial plane, and the time-averaged intensity is diffraction-free [Fig.~\ref{fig:MonochrBeamTransformation}(f)].

\section{Lorentz boost of space-time wave packets}\label{section:transfBaseband}

In this Section, we show that the Lorentz boost of an STWP is another STWP. Therefore, the result that was demonstrated in the previous Section (that the iso-frequency plane associated with a monochromatic beam in $\mathcal{O}$ is tilted upon a Lorentz boost in $\mathcal{O}'$) can be generalized: the tilted spectral plane associated with an STWP in $\mathcal{O}$ is tilted upon a Lorentz boost in $\mathcal{O}'$. However, by maintaining $|\beta|\!<\!1$, there are restrictions on the amount of tilt that can be introduced into the spectral plane. This leads to a natural separation between the subluminal and the superluminal regimes of STWPs. 

\begin{figure}[t!]
    \centering
    \includegraphics[width=8.6cm]{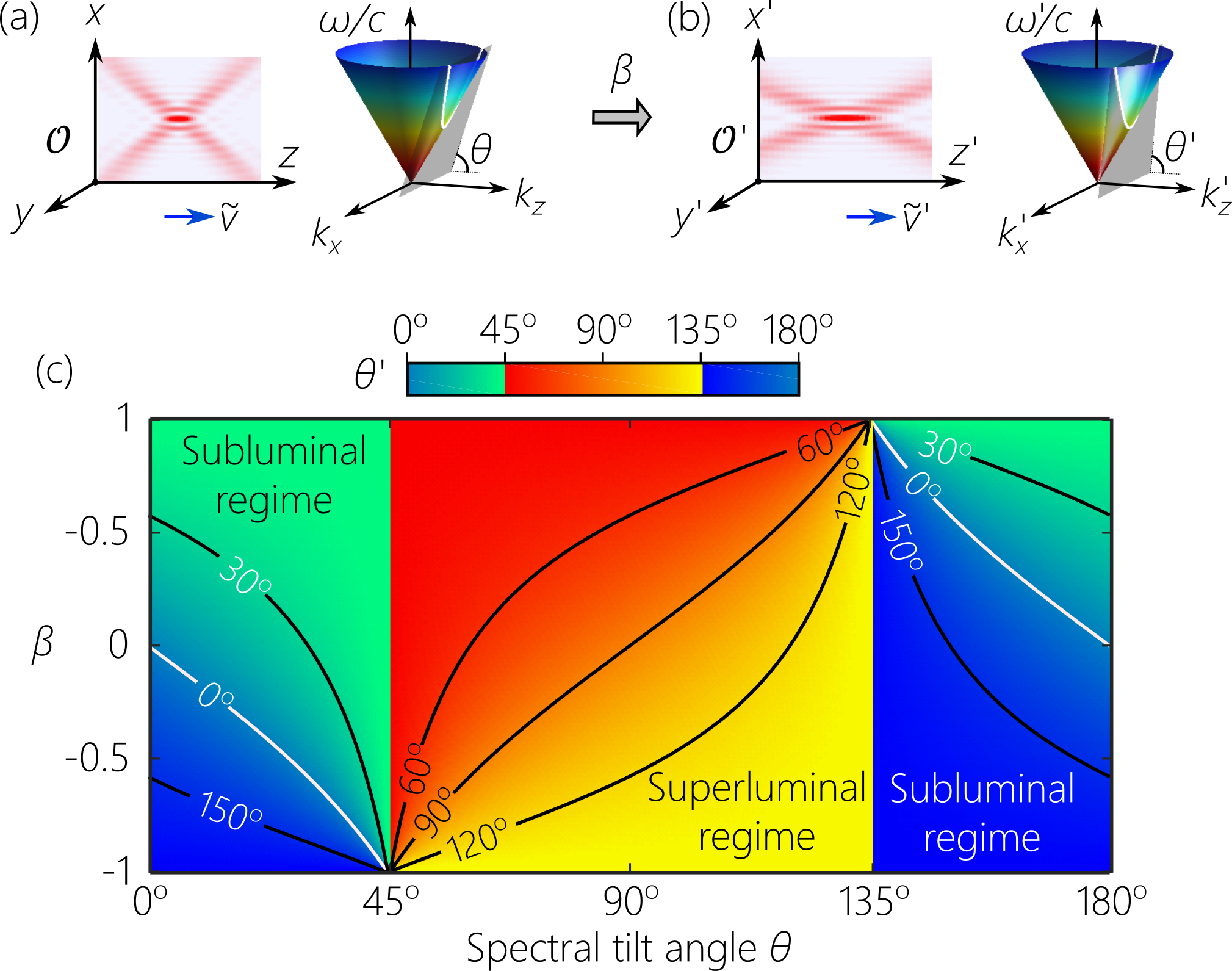}
    \caption{(a) We start with an STWP in the rest frame $\mathcal{O}$ characterized by a spectral tilt angle $\theta$ and thus a group velocity $\widetilde{v}\!=\!c\tan{\theta}$. (b) In the moving frame $\mathcal{O}'$, the Lorentz boost transforms the STWP in $\mathcal{O}$ to another STWP characterized by the spectral tilt angle $\theta'$. The spatial bandwidth $\Delta k_{x}\!=\!\Delta k_{x}'$ is invariant, but the temporal bandwidth $\Delta\omega$ and the carrier frequency $\omega_{\mathrm{o}}$ change. (c) Plot of the spectral tilt angle $\theta'$ in $\mathcal{O}'$ as a function of the spectral tilt angle $\theta$ in $\mathcal{O}$ and the Lorentz parameter $\beta$. We use two different color palettes to demarcate the subluminal and superluminal regimes. The dashed white curves identify the locus of $\tan\theta\!=\!\beta$.}
    \label{fig:PulsedBeamTransformation}
\end{figure}

By taking as our starting point an STWP in the rest frame $\mathcal{O}$ that corresponds to a spectral tilt angle $\theta$, it is straightforward to show that a Lorentz boost (Eq.~\ref{Eq:LorentzTransFourierDomain}) to the moving frame $\mathcal{O}'$ changes the spectral tilt angle to $\theta'$ [Fig.~\ref{fig:PulsedBeamTransformation}(a,b)], such that:
\begin{equation}
\omega-\omega_{\mathrm{o}}\!=\!(k_{z}-k_{\mathrm{o}})c\tan{\theta}\Longrightarrow\omega'-\omega_{\mathrm{o}}'\!=\!(k_{z}'-k_{\mathrm{o}}')c\tan{\theta'},
\end{equation}
where $k_{\mathrm{o}}\!=\!\tfrac{\omega_{\mathrm{o}}}{c}$, $\omega_{\mathrm{o}}'\!=\!\omega_{\mathrm{o}}\gamma(1-\beta)$, $k_{\mathrm{o}}'\!=\!\tfrac{\omega_{\mathrm{o}}'}{c}$, and the new spectral tilt angle $\theta'$ is related to the initial angle $\theta$ via:
\begin{equation}\label{Eq:TransformingTheta}
\tan{\theta'}=\frac{\tan\theta-\beta}{1-\beta\tan\theta}.
\end{equation}
We plot $\theta'$ in Fig.~\ref{fig:PulsedBeamTransformation}(c) as a function of $\theta$ and $\beta$. Expressed in terms of velocities, this transformation takes the form:
\begin{equation}\label{Eq:TransformingV}
\widetilde{v}'=\frac{\widetilde{v}-v}{1-\frac{\widetilde{v}v}{c^{2}}},
\end{equation}
where $\widetilde{v}\!=\!c\tan{\theta}$ is the group velocity of the STWP in the rest frame $\mathcal{O}$, $\widetilde{v}'\!=\!c\tan{\theta'}$ is the group velocity of the STWP in the moving frame $\mathcal{O}'$, and $v$ is the relative velocity between $\mathcal{O}'$ and $\mathcal{O}$. Note that this formula is nothing but the usual relativistic formula for combining velocities typically associated with massive particles. In other words, when implementing Lorentz boosts to STWPs, these wave packets can be regarded to act as massive particles with respect to their group velocity. Note that we retrieve the result for the monochromatic beam presented in the previous Section by setting $\theta\!=\!0$ in $\mathcal{O}$, thereby yielding $\widetilde{v}'\!=\!-v$ ($\tan\theta'\!=\!-\beta$).

Although we assume that $|\beta|\!<\!1$, we need not put any restrictions on $\theta$. This result shows, in general, that the Lorentz boost of an STWP is another STWP: the Lorentz boost of a subluminal STWP in $\mathcal{O}$ is another subluminal STWP in $\mathcal{O}'$, and the Lorentz boost of a superluminal STWP in $\mathcal{O}$, on the other hand, is another superluminal STWP in $\mathcal{O}'$. This is clear from the plot of $\theta'$ in Fig.~\ref{fig:PulsedBeamTransformation}(c).

\subsection{Subluminal regime for STWPs}\label{section:transfSublumBaseband}

\begin{figure}[t!]
    \centering
    \includegraphics[width=8.6cm]{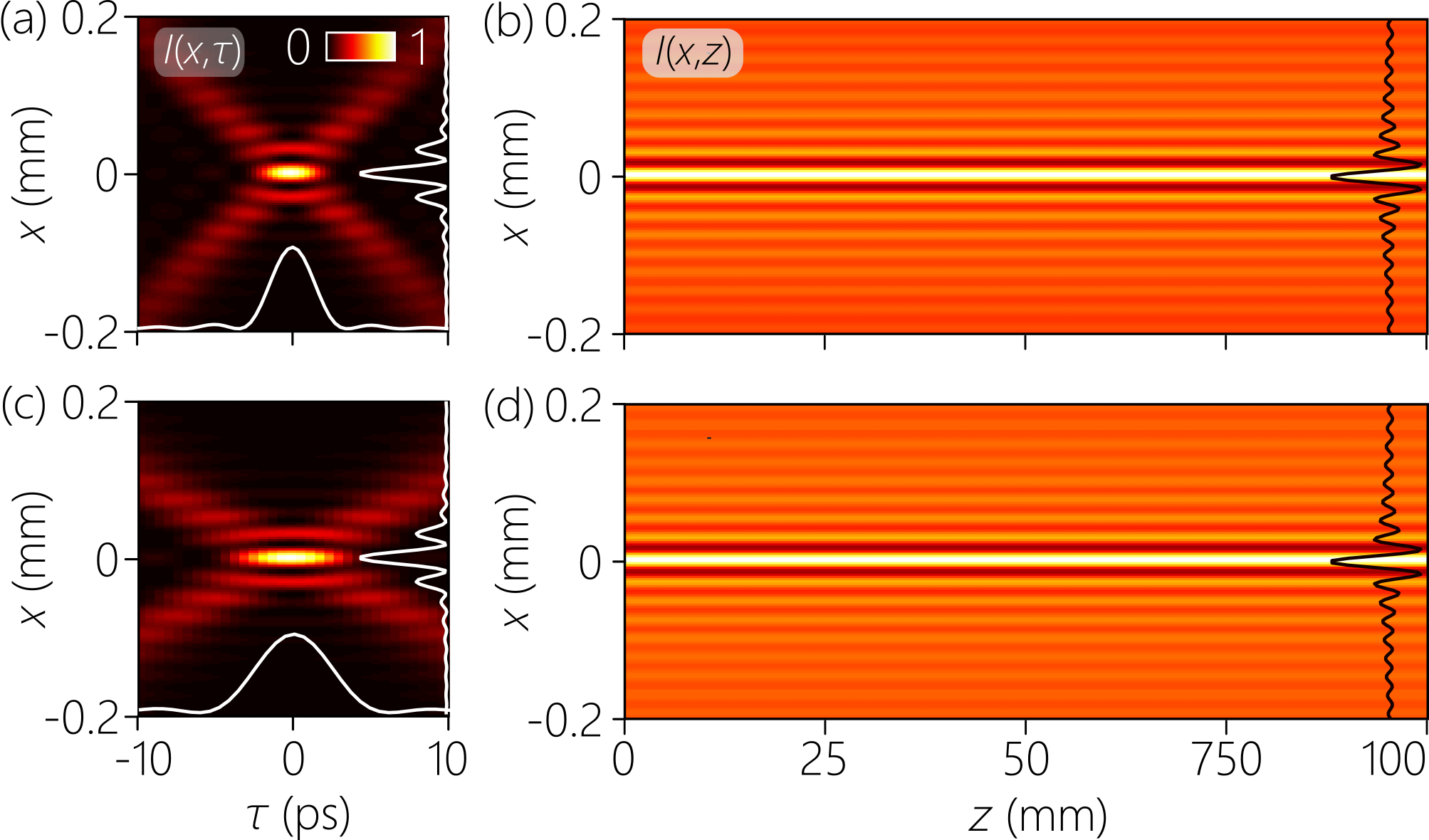}
    \caption{(a) The spatio-temporal intensity profile $I(x,0;t)$ at the axial plane $z\!=\!0$ in the rest frame $\mathcal{O}$ for an STWP with $\theta\!=\!40^{\circ}$; $\Delta k_{x}\!=\!0.17$~rad/$\mathrm{\mu m}$ and $\Delta\omega\!=\!2\pi\times0.5$~THz. The white curves correspond to the temporal profile at the beam center $I(0,0;t)$, and the spatial beam profile at the pulse center $I(x,0;0)$. (b) The time-averaged intensity $I(x,z)$ for the STWP in (a). (c) Same as (a) after implementing the Lorentz boost associated with the moving frame $\mathcal{O}'$. Note that the spatial beam profile at the pulse profile has not changed. (d) The time-averaged intensity $I(x,z)$ for the STWP in (c).}
    \label{fig:EnvelopeTransformation}
\end{figure}

The results described above, in particular, Eq.~\ref{Eq:TransformingTheta} and Fig.~\ref{fig:PulsedBeamTransformation}(c), show that the Lorentz boost of a subluminal STWP in $\mathcal{O}$ is another subluminal STWP in $\mathcal{O}'$. The transformation of the field from $E(x,z;t)$ in $\mathcal{O}$ to $E'(x',z';t')$ in $\mathcal{O}'$ follows Eq.~\ref{Eq:GeneralFieldTransformation}. We plot in Fig.~\ref{fig:EnvelopeTransformation}(a) the spatio-temporal intensity profile $I(x,z;t)$ for a subluminal STWP in $\mathcal{O}$ with $\theta\!=\!40^{\circ}$, and plot in Fig.~\ref{fig:EnvelopeTransformation}(c) the corresponding intensity profile $I'(x',z';t')$ in $\mathcal{O}'$ when $\beta\!=\!0.4$, whereupon $\theta'\!=\!33.5^{\circ}$. Because the spatial frequencies remain invariant upon the Lorentz boost $k_{x}'\!=\!k_{x}$, the spatial beam profile at the pulse center at any fixed axial plane is therefore also invariant $I(x,0;0)\!=\!I'(x,0;0)$, at $z\!=\!z'\!=\!0$ for example.

On the other hand, the temporal bandwidth $\Delta\omega_{1}$ of a subluminal STWP with group velocity $\widetilde{v}_{1}\!=\!c\tan\theta_{1}$ is related to the temporal bandwidth $\Delta\omega_{2}$ of another subluminal STWP with group velocity $\widetilde{v}_{2}\!=\!c\tan\theta_{2}$ that shares the \textit{same spatial bandwidth} (and therefore can be related via a Lorentz boost) via: 
\begin{equation}\label{Eq:ChangeInBandwdith}
\frac{\Delta\omega_{2}}{\Delta\omega_{1}}=\frac{\gamma_{2}}{\gamma_1}\left|\frac{\widetilde{v}_{2}}{\widetilde{v}_{1}}\right|,
\end{equation}
assuming both STWPs are in the paraxial domain, and $\theta_{1}\!\neq\!0^{\circ}$; here $\gamma_{1}\!=\!1/\sqrt{1-(\widetilde{v}_{1}/c)^{2}}$, and $\gamma_{2}\!=\!1/\sqrt{1-(\widetilde{v}_{2}/c)^{2}}$. If we take the STWP with group velocity $\widetilde{v}_{1}$ to be in $\mathcal{O}$ and the other in $\mathcal{O}'$, then the relative velocity between these two frames is $\beta\!=\!\tfrac{\tan\theta_{1}-\tan\theta_{2}}{1-\tan\theta_{1}\tan\theta_{2}}$. For the example in Fig.~\ref{fig:EnvelopeTransformation}; because $\widetilde{v}_{1}\!=\!0.84c$ and $\widetilde{v}_{2}\!=\!0.66c$, the ratio between the temporal bandwidths is $\tfrac{\Delta\omega_{2}}{\Delta\omega_{1}}\!\approx\!0.57$. This can be clearly seen in the change in the temporal width of the pulse profile at the beam center $I(0,0;t)$ in $\mathcal{O}$ [Fig.~\ref{fig:EnvelopeTransformation}(a)] compared to that of $I'(0,0;t')$ [Fig.~\ref{fig:EnvelopeTransformation}(c)]. The time-averaged intensity of either STWP is independent of the temporal bandwidth and depends only on their spatial bandwidth. Because the spatial bandwidth is invariant under a Lorentz boost, the $I(x,z)$ is identical for both STWPs [Fig.~\ref{fig:EnvelopeTransformation}(b,d)].

We can draw several general conclusions from the formula in Eq.~\ref{Eq:ChangeInBandwdith} for the change in temporal bandwidth upon a Lorentz boost. First, when $\beta\!=\!-\tan\theta_{1}$, then $\theta_{2}\!=\!0^{\circ}$; that is, the finite temporal bandwidth of the STWP is reduced to zero, $\Delta\omega_{2}\!=\!0$ and the STWP reverts to a monochromatic beam, corresponding to the white curves in Fig.~\ref{fig:PulsedBeamTransformation}(c). Second, the temporal bandwidth remains invariant $\Delta\omega_{1}\!=\!\Delta\omega_{2}$ in two cases: the first is the trivial case when $\beta\!=\!0$, whereupon $\widetilde{v}_{1}\!=\!\widetilde{v}_{2}$ and $\gamma_{1}\!=\!\gamma_{2}$; and the second occurs when $\widetilde{v}_{2}\!=\!-\widetilde{v}_{1}$ ($\gamma_{1}\!=\!\gamma_{2}$), whereupon $\beta\!=\!\tfrac{2\tan\theta_{1}}{1+\tan^{2}\theta_{1}}$ \cite{Saari20JPC}.



\begin{figure*}[t!]
    \centering
    \includegraphics[width=17.6cm]{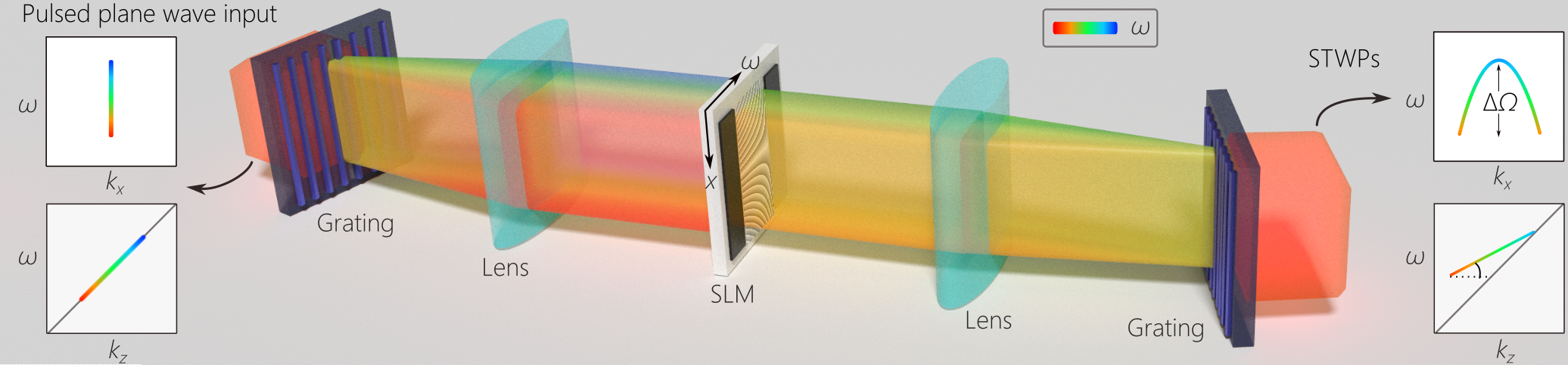}
    \caption{Overall setup for synthesizing STWPs and implementing Lorentz boosts. For clarity, the setup is shown unfolded with the SLM depicted in transmission mode. In our experiment, we made use of a reflective SLM, so that the retro-reflected field returns to the same initial grating. The transverse width of the SLM phase pattern determines the value of the factor $\beta$. Left inset: spatio-temporal spectrum of the initial plane-wave laser pulse projected onto the $(k_{x},\tfrac{\omega}{c})$ and $(k_{z},\tfrac{\omega}{c})$ planes. Right inset: same as left inset for the synthesized STWP. The dashed lines are the light-lines $k_{z}\!=\!\tfrac{\omega}{c}$.}
    \label{fig:Setup}
\end{figure*}

\subsection{Superluminal group velocities}\label{section:transfSuperlumBaseband}

According to Eq.~\ref{Eq:TransformingTheta} and Eq.~\ref{Eq:TransformingV}, a monochromatic beam or a subluminal STWP in $\mathcal{O}$ is transformed into either a monochromatic beam or a subluminal STWP in $\mathcal{O}'$ when $|\beta|\!<\!1$. However, it is by now well-established that superluminal STWPs exist, which simply implies varying the association between spatial and temporal frequencies as determined by Eq.~\ref{eq:SpectralCorrelation} in the range when $45^{\circ}\!<\!\theta\!<\!135^{\circ}$ for the spectral tilt angle, and does \textit{not} imply any violation of relativistic causality \cite{SaariPRA18,Yessenov22AOP}. Indeed, Eq.~\ref{Eq:TransformingTheta} implies that a superluminal STWP in $\mathcal{O}$ is transformed into another superluminal STWP in $\mathcal{O}'$. The change in the group velocity for superluminal STWPs can be seen in the middle section of Fig.~\ref{fig:PulsedBeamTransformation}(c). In this case, Eq.~\ref{Eq:ChangeInBandwdith} still applies as long as $\theta_{1},\theta_{2}\!\neq\!90^{\circ}$. Of course, there is no Lorentz boost that can transform a superluminal STWP into a monochromatic beam.

\section{Experimental demonstration}

\begin{figure*}[t!]
    \centering
    \includegraphics[width=17cm]{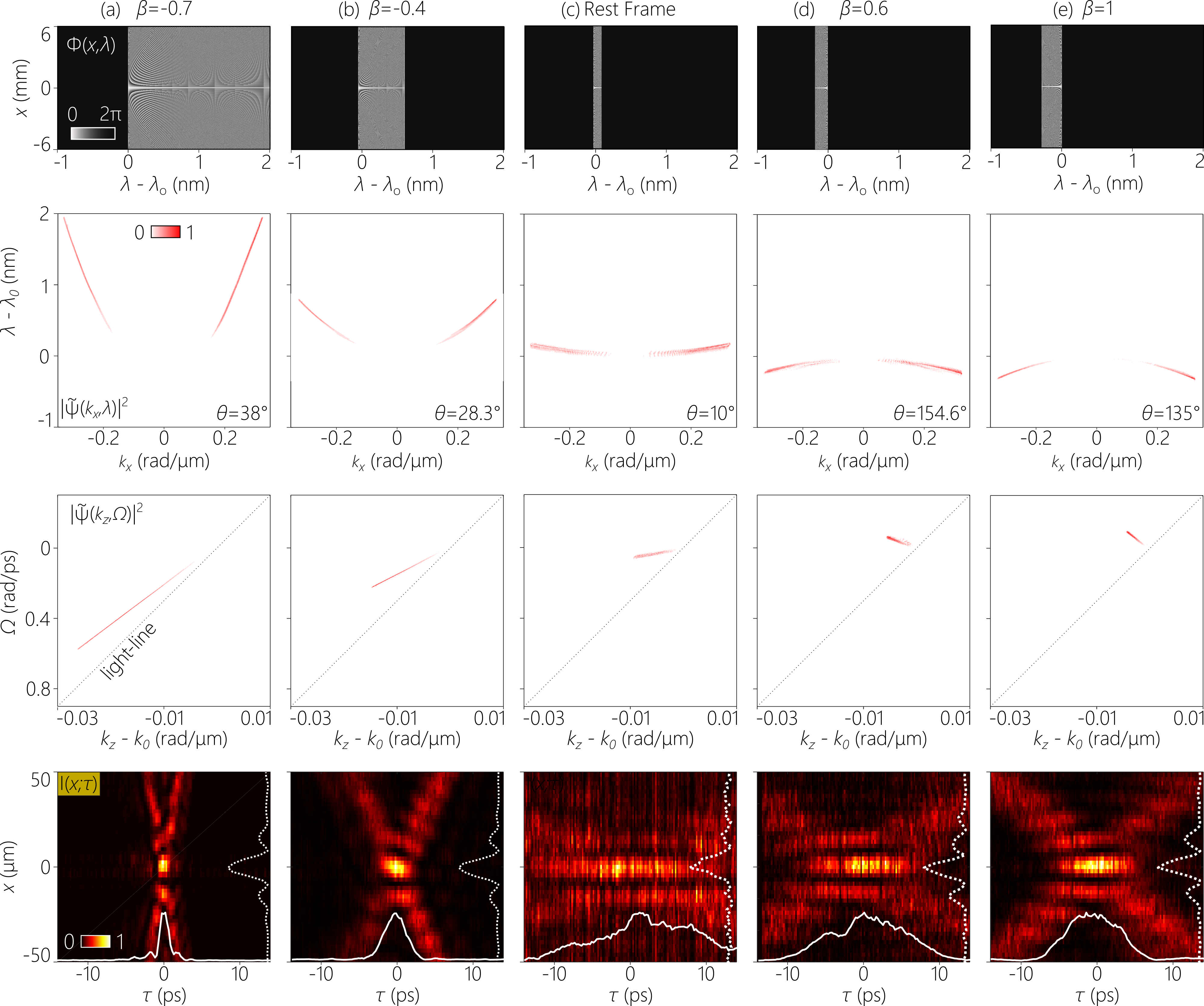}
    \caption{Experimental realization of Lorentz boosts of subluminal STWPs. Rows from top to bottom: first row is the SLM phase pattern $\Phi$ to produce the STWP; second row, the measured spatio-temporal spectrum projected onto the $(k_{x},\lambda)$-plane; third row, the spatio-temporal spectrum from the second row projected onto the $(k_{z},\Omega)$-plane; and fourth row, measured spatio-temporal intensity profile $I(x,z;\tau)$ at a fixed axial plane $z$. The dark portions of the phase distributions correspond to spectral components that are not spatially modulated, and are subsequently eliminated by a Fourier plane spatial filter. The dashed line in the $(k_{z},\Omega)$-plane is the light-line $k_{z}\!=\!\tfrac{\omega}{c}$. (a) Subluminal STWP obtained by a Lorentz boost with $\beta\!=\!-0.7$, (b) $\beta\!=\!-0.4$, (d) $\beta\!=\!0.6$, and (e) $\beta\!=\!1$, starting from the STWP in the rest frame shown in (c). Note the change in the sign of the curvature of the spatio-temporal spectrum in the $(k_{x},\lambda)$-plane for positive and negative values of $\beta$.}
    \label{fig:DataSublumST}
\end{figure*}

A Lorentz boost modifies an STWP as follows: (1) the temporal bandwidth changes according to Eq.~\ref{Eq:ChangeInBandwdith}; (2) the beam width at the pulse center $\Delta x$ is invariant; (3) the spatio-temporal intensity profile changes accordingly; (4) the group velocity changes according to Eq.~\ref{Eq:TransformingTheta}; and (5) the carrier frequency is Doppler-shifted, without impacting the spatio-temporal intensity profile in the narrowband paraxial regime. In this Section, we present a table-top experimental approach that emulates consequences (1) through (4) of a Lorentz boost, but not the carrier Doppler shift in (5). We first describe the overall optical setup utilized [Fig.~\ref{fig:Setup}], and then explain the methodology for implementing a Lorentz boost [Fig.~\ref{fig:SetupBetaControl}], before presenting our experimental results in the subluminal [Fig.~\ref{fig:DataSublumST}] and superluminal [Fig.~\ref{fig:DataSuperlumST}] regimes, and measurements of group velocity change in these regimes [Fig.~\ref{fig:DataGroupVelocity}].

\subsection{Methodology of the Lorentz transformation of STWPs}

We make use of the optical system depicted in Fig.~\ref{fig:Setup} for the synthesis of STWPs \cite{Kondakci17NP,Yessenov22AOP,Yessenov20PRL2}. We start with 100-fs pulses from a mode-locked Ti:Sapphire laser (Tsunami, Spectra physics; spectral bandwidth $\Delta\lambda\approx10$~nm centered at $\lambda\approx800$~nm). We expand the laser beam to approximate a plane-wave pulse before directing it to a diffraction grating (Newport 10HG1200-800-1, 1200 lines/mm), whereupon the first diffraction order is selected and collimated via a cylindrical lens ($f\!=\!50$~cm) in a $2f$ configuration. The pulse spectrum is spatially resolved at the focal plane of the lens, where we place a 2D reflective, phase-only spatial light modulator (SLM, Hamamatsu X10468-02), which imparts a phase profile to the incident field. The temporal spectrum is spread spatially along one dimension of the SLM active area. With the given grating and lens, the SLM intercepts only a bandwidth of $\Delta\lambda\!\approx\!2$~nm of the incident spectrum. The maximum temporal bandwidth of any synthesized STWP here is $\Delta\lambda\!\approx\!2$~nm. After modulation by the SLM phase, the field is retro-reflected back through the cylindrical lens to the grating whereupon the spectrum is recombined and the STWP constituted. We depict this configuration unfolded in Fig.~\ref{fig:Setup} for clarity, with the SLM in transmission rather than reflection mode. 

\begin{figure*}[t!]
    \centering
    \includegraphics[width=16cm]{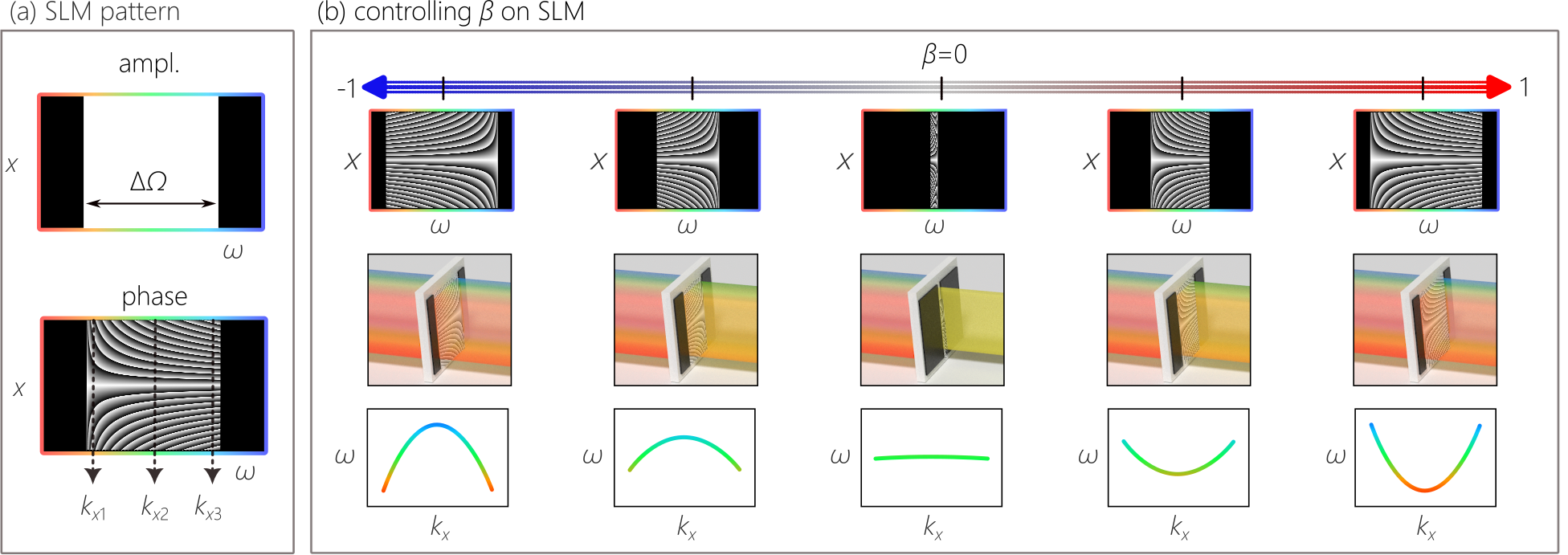}
    \caption{(a) A temporal spectrum of bandwidth $\Delta\omega$ is intercepted by the SLM (top), and a 2D spatial phase distribution (bottom) is imparted to the spectrally resolved wavefront. (b) Change in the phase distribution (top row), the concomitant change in the temporal bandwidth transmitted through the optical system (middle row), and the concomitant change in the spatio-temporal spectrum projected onto the $(k_{x},\omega)$-plane associated with Lorentz boosts indexed by $\beta$. The scenario depicted is that of the transformation of a monochromatic optical beam (center column, $\beta\!=\!0$) into a subluminal STWP. Note the change in the sign of the curvature of the spatio-temporal spectrum for positive and negative values of $\beta$, which results in a change in the sign of the STWP group velocity.}
    \label{fig:SetupBetaControl}
\end{figure*}

We characterize the STWPs in the spectral and physical domains. We measure $|\widetilde{\psi}(k_{x},\lambda)|^{2}$ in the spectral domain by implementing a spatio-temporal Fourier transform (the spatial Fourier transform using a lens in a $2f$ configuration, and the temporal Fourier transform using a grating). This spectrum allows us to estimate the spectral tilt angle $\theta$ and thus the group velocity $\widetilde{v}\!=\!c\tan{\theta}$. In the physical domain, the spatio-temporal intensity profile $I(x,z;\tau)$ at a fixed axial plane $z$ is obtained by placing the entire synthesis setup as depicted in Fig.~\ref{fig:Setup} in one arm of a Mach-Zehnder interferometer, with an optical delay $\tau$ in the other arm, which is thus traversed by the initial 100-fs laser pulses as a reference. While sweeping the delay $\tau$ in the reference arm, we monitor the visibility of the spatially resolved interference fringes resulting from the interference of the STWP (of maximum temporal bandwidth $\Delta\lambda\!\leq\!2$~nm) with the 100-fs ($\Delta\lambda\!\approx\!10$~nm) reference pulse. To measure the group velocity of an STWP, we repeat the measurement of the spatio-temporal intensity profile $I(x,z;\tau)$ at a different axial plane $z$ and from the relative change in the delay line $c\Delta\tau$ and the camera position $\Delta z$ we calculate the group velocity of the STWP in free space; see Refs.~\cite{Kondakci19NC,Yessenov19OE} for details. 

In our experiments, a fixed distribution of wavelengths is delivered across the SLM and maintained in all the measurements [Fig.~\ref{fig:SetupBetaControl}(a)]. By changing the 2D spatial phase distribution $\Phi$ imparted, we can implement a Lorentz boost with any prescribed value of $\beta$ (within the limit of maximum available bandwidth $\Delta\lambda\!\approx\!2$~nm). Each wavelength occupies a column of the SLM. The SLM phase distribution associated with each frequency $\omega$ is linear $\Phi(x)\!=\!k_{x}x$, where $x$ is the coordinate perpendicular to the direction of the spread spectrum, and $k_{x}(\omega)$ is selected according to Eq.~\ref{eq:SpectralCorrelation} [Fig.~\ref{fig:SetupBetaControl}(a)]. When assigned to the temporal frequency $\omega$, this phase is $\Phi(\omega,x)\!=\!(\tfrac{\omega}{c}\sin\varphi)x$, where $\varphi$ is the propagation angle of this spatial frequency with the $z$-axis.

We consider a spatial bandwidth $\Delta k_{x}$ that is invariant under any Lorentz boost. The SLM thus implements spatial frequencies extending from $k_{x}\!=\!0$ to $k_{x}\!=\!\Delta k_{x}$. We implement SLM phase distributions with mirror symmetry along the vertical $x$-axis around its center to incorporate positive and negative spatial frequencies $\pm k_{x}(\omega)$ with each frequency $\omega$ (to produce a beam that is spatially symmetric along $x$). As we vary $\beta$, the arrangement of this fixed set of spatial frequencies is modified across the SLM. When exploiting the full temporal bandwidth $\Delta\lambda\!\approx\!2$~nm, the spatial frequencies are spread across the entire width of the SLM. Under a Lorentz boost, the temporal bandwidth changes (Eq.~\ref{Eq:ChangeInBandwdith}). To capture this effect, the fixed collection of spatial frequencies is either compressed or stretched across the SLM active area. Consequently, some wavelengths are not modulated by any spatial frequency. We set the SLM phase for these frequencies to $\Phi(\omega,x)\!=\!0$, and then place in the path of the synthesized STWP a $4f$ imaging system with a spatial beam block in its Fourier plane. This beam block removes any part of the field in the vicinity of $k_{x}\!=\!0$ and thus eliminates all wavelengths from the spectrum that were not spatially modulated by the SLM.

Starting with a monochromatic beam, where all the spatial frequencies are imparted to a single frequency $\omega_{\mathrm{o}}$ [Fig.~\ref{fig:SetupBetaControl}(b), $\beta\!=\!0$], positive- or negative-valued $\beta$ produces subluminal STWPs with negative ($135^{\circ}\!<\!\theta\!<\!180^{\circ}$) or positive ($0^{\circ}\!<\!\theta\!<\!45^{\circ}$) group velocity, respectively. This is captured by changing the direction of ordering of the spatial frequencies from $k_{x}\!=\!0$ to $\Delta k_{x}$ or from $k_{x}\!=\!\Delta k_{x}$ down to 0 [Fig.~\ref{fig:SetupBetaControl}(b) to the left and right of $\beta\!=\!0$]. The spatial frequency $k_{x}\!=\!0$ is associated with the frequency $\omega_{\mathrm{o}}$, so that the two different orientations of the phase distribution is associated with the frequencies $\omega_{\mathrm{o}}$ (assigned to $k_{x}\!=\!0$) to $\omega_{\mathrm{o}}-\Delta\omega$ (assigned to $k_{x}\!=\!\Delta k_{x}$), or -- in the opposite orientation -- the frequencies $\omega_{\mathrm{o}}+\Delta\omega$ (assigned to $k_{x}\!=\!0$) to $\omega_{\mathrm{o}}$ (assigned to $k_{x}\!=\!\Delta k_{x}$). The spectral content that is \textit{not} spatially modulated [the dark area in Fig.~\ref{fig:SetupBetaControl}(b)] is filtered out by the spatial filter. Implementing a Lorentz boost with larger $|\beta|$ broadens the temporal bandwidth, which is achieved by spreading the collection of spatial frequencies over the SLM width. The spatially unmodulated spectral content drops and the temporal bandwidth of the STWP increases.

\subsection{Measurements for subluminal STWPs}\label{section:ExpSublum}

\begin{figure*}[t!]
    \centering
    \includegraphics[width=17cm]{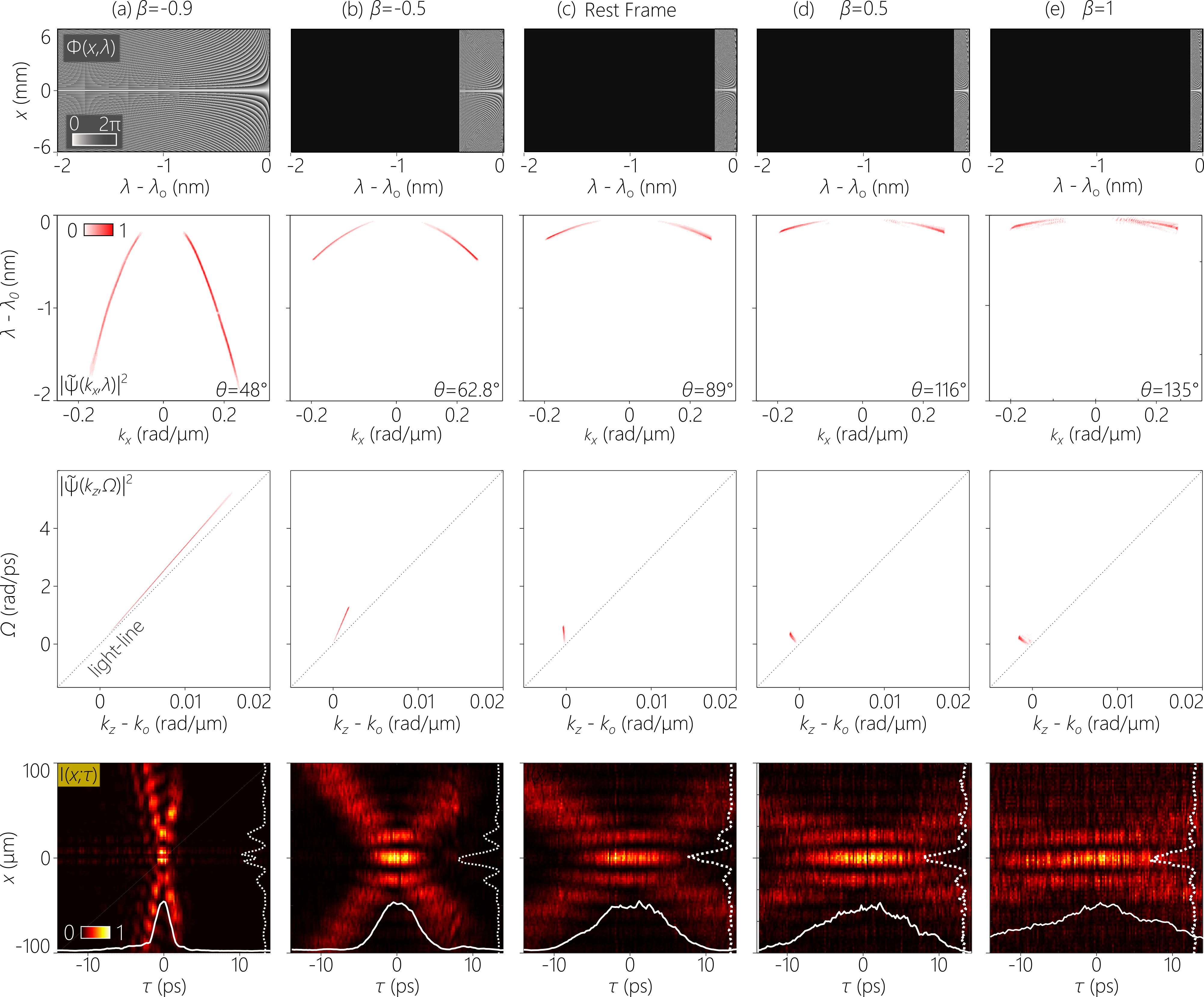}
    \caption{Experimental realization of Lorentz boosts of superluminal STWPs. Rows from top to bottom are the same as in Fig.~\ref{fig:DataSublumST}. (a) Superluminal STWP obtained by a Lorentz boost with $\beta\!=\!-0.9$, (b) $\beta\!=\!-0.5$, (d) $\beta\!=\!0.5$, and (e) $\beta\!=\!1$, starting from the STWP in the rest frame shown in (c). Note that the sign of the curvature of the spatio-temporal spectrum in the $(k_{x},\lambda)$-plane is the same for both positive and negative values of $\beta$.}
    \label{fig:DataSuperlumST}
\end{figure*}

To implement a monochromatic beam in the rest frame $\mathcal{O}$ in our setup requires imparting all the spatial frequencies from $k_{x}\!=\!0$ to $\Delta k_{x}$ to a single temporal frequency, which corresponds to a single SLM column [$\beta\!=\!0$ in Fig.~\ref{fig:SetupBetaControl}(b)]. Because this cannot be done in practice, we take as the starting point in the subluminal regime a quasi-monochromatic beam. This is realized by compressing all the spatial frequencies over a narrow extent $\Delta\lambda$ along the direction of the spread spectrum [Fig.~\ref{fig:DataSublumST}(c)], such that the spectral projection onto the $(k_{x},\lambda)$-plane approaches a horizontal line. Here, the pulse width at the beam center $I(0;\tau)$ is wide, and the beam width at the pulse center $I(x;0)$ is proportional to the inverse of the spatial bandwidth, $\Delta x\!\sim\!\tfrac{1}{\Delta k_{x}}$. The spatial frequencies are compressed over a fraction of the SLM active area, with only $\Delta\lambda\!\approx\!0.14$~nm spatially modulated by the SLM. The rest of the $\Delta\lambda\!\approx\!2$~nm bandwidth incident on the SLM is \textit{not} spatially modulated (assigned $k_{x}\!=\!0$). The Fourier spatial filter placed in the path of the STWP eliminates the spatial frequencies in the vicinity of $k_{x}\!=\!0$, thereby passing only the spatially modulated portion of the spectrum.

Positive values of $\beta$ result in negative group velocities of the produced STWP, and the associated spectral tilt angles are in the range $135^{\circ}\!<\!\theta\!<\!180^{\circ}$. The case of $\beta\!=\!0.6$ is shown in Fig.~\ref{fig:DataSublumST}(d), corresponding to $\theta\!=\!154.6^{\circ}$. The projected spectrum in this case is a segment of an ellipse. Note that $\beta\!=\!1$ can be realized, whereupon $\theta\!=\!135^{\circ}$ and $\widetilde{v}\!=\!-c$ [Fig.~\ref{fig:DataSublumST}(e)]. The corresponding conic section on the light-cone is a parabola.

Negative values of $\beta$, on the other hand, result in positive group velocities of the produced STWP, and the associated spectral tilt angles are in the range $0^{\circ}\!<\!\theta\!<\!45^{\circ}$. This regime requires flipping the orientation of the arrangement of spatial frequencies. The case of $\beta\!=\!-0.4$ is shown in Fig.~\ref{fig:DataSublumST}(b), corresponding to $\theta\!=\!28.3^{\circ}$. The maximum value of $\beta$ that can be realized here is $\beta\!=\!-0.7$ as shown in Fig.~\ref{fig:DataSublumST}(a). In all cases [Fig.~\ref{fig:DataSublumST}(a,b,d,e)], the temporal bandwidth $\Delta\omega_{2}$ is related to $\Delta\omega_{1}$ of the STWP in Fig.~\ref{fig:DataSublumST}(c) through Eq.~\ref{Eq:ChangeInBandwdith}.

\subsection{Measurements for superluminal STWPs}

We plot in Fig.~\ref{fig:DataSuperlumST} the measurement results of Lorentz boosts in the superluminal regime. In this case, we take as our starting point in the rest frame $\mathcal{O}$ a superluminal STWP corresponding to $\theta\!=\!89^{\circ}$ [Fig.~\ref{fig:DataSuperlumST}(c)]; i.e., $\widetilde{v}\!\approx\!60c$. We implement Lorentz boosts to $\mathcal{O}'$ corresponding to negative-valued [Fig.~\ref{fig:DataSuperlumST}(a,b)] and positive-valued [Fig.~\ref{fig:DataSuperlumST}(d,e)] $\beta$. In this superluminal scenario, the sign of the curvature of the spatiotemporal spectrum projected onto the $(k_{x},\lambda)$-plane remains the same for all values of $\beta$ [Fig.~\ref{fig:DataSuperlumST}, second row]. Therefore, the orientation of the arrangement of spatial frequencies remains the same. The temporal bandwidth $\Delta\omega_2$ drops with respect to $\Delta\omega_1$ for $\beta\!>\!0$, resulting in a broadening of the temporal width of the pulse, whereas $\Delta\omega_2$ increases when $\beta\!<\!0$ [Fig.~\ref{fig:DataSuperlumST}, fourth row]. Once again, the change in the temporal bandwidth with $\beta$ follows the relationship in Eq.~\ref{Eq:ChangeInBandwdith}.

\subsection{Measurements of the group velocity change}
\begin{figure}[t!]
    \centering
    \includegraphics[width=8.5cm]{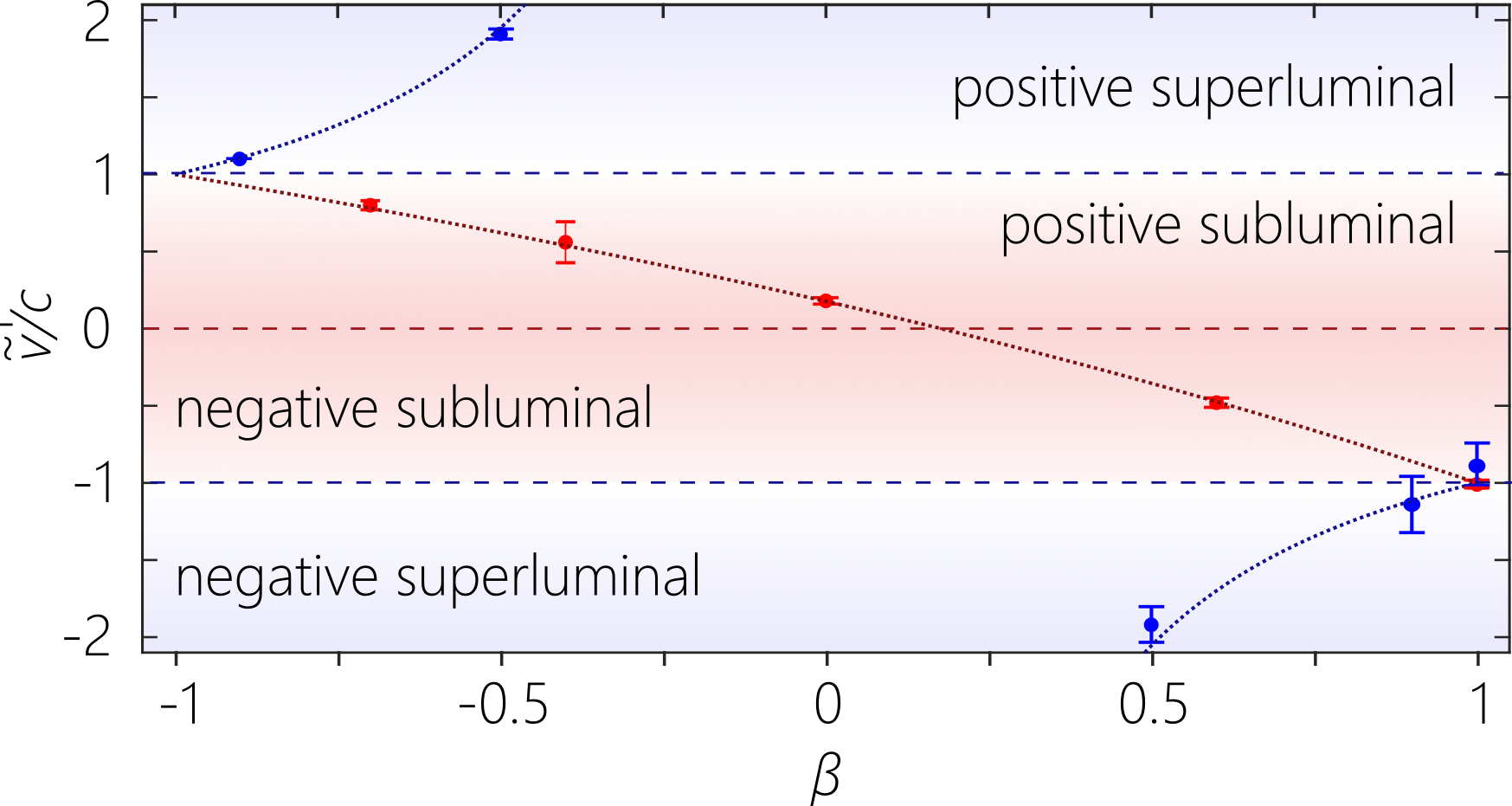}
    \caption{Measured group velocities of STWPs after Lorentz boost corresponding to the measurements in Fig.~\ref{fig:DataSublumST} and Fig.~\ref{fig:DataSuperlumST}. Dots indicate the measured data with error bars, dotted lines correspond to the theoretical expectations from Eq.~\ref{Eq:TransformingV} and dashed lines show the boundaries between the subluminal (red region) and superluminal (blue region) regimes.}
    \label{fig:DataGroupVelocity}
\end{figure}

We plot in Fig.~\ref{fig:DataGroupVelocity} measured group velocities of subluminal and superluminal STWPs characterized in Fig.~\ref{fig:DataSublumST} and Fig.~\ref{fig:DataSuperlumST}. Starting with the STWP in the rest frame ($\beta\!=\!0$) traveling at $\widetilde{v}\!=\!0.18c$ in the subluminal regime [Fig.~\ref{fig:DataSublumST} (c)] and $\widetilde{v}\!=\!62c$ in the superluminal regime [Fig.~\ref{fig:DataSuperlumST} (c)], we confirm that the group velocities of the STWPs in the moving frames follow the relativistic velocity addition formula in Eq.~\ref{Eq:TransformingV} (dotted lines). In addition, these measurements support that STWPs do not cross the subluminal-superluminal barrier (dashed lines) after a Lorentz boost. 

\section{Discussion and conclusions}

We have couched our results in terms of coherent fields, whether monochromatic plane waves, monochromatic paraxial coherent beams, or coherent STWPs. Nevertheless, our experimental approach is capable of handling incoherent fields, as demonstrated in \cite{Yessenov19Optica,Yessenov19OL}. This should motivate extending the theoretical formulation of Lorentz boosts to partially coherent and incoherent optical fields, where almost no previous work has been done. Moreover, studying the impact of Lorentz boosts on vector fields \cite{Guo21Light,Diouf21OE,Yessenov22OL} would be of interest.

Recent work has been devoted to the study of light-matter interaction involving moving photonic devices \cite{Leger2019AP,Caloz2020IEEE1,Caloz2020IEEE2,Rocca2020IEEE}, such as gratings \cite{Kowalski06PLA,Bahabad14OQE}, Bragg mirrors, photonic crystals, and waveguides \cite{Qu2016JOSAB}. The interaction of STWPs and their coupling to waveguides \cite{Shiri20NC,Guo21PRR,Bejot21ACSP,Shiri22ACSP,Shiri23JOSAA,Jolly23JO,Stefanska23ACSP} and planar Fabry-P{\'e}rot cavities \cite{Shiri20OL,Shiri20APLP} has recently made significant strides. The work presented here regarding the Lorentz boost of STWPs can help connect these new results with the theoretical work on the interaction of light with moving devices. Our work may also benefit the study of photonic time crystals \cite{Lyubarov2022Science,Sharabi2022Optica}, the realizations of optical analogs of the Mackinnon wave packet \cite{Mackinnon78FP} via moving dipoles \cite{WilczekBook,Hall23NatPhys}; and the reflection and refraction of optical waves from moving interfaces \cite{Plansinis15PRL,Qu2016JOSAB,Plansinis16JOSAB,Plansinis17JOSAB,Plansinis18JOSAB,Leger2021Photonics}.

In conclusion, we have presented a theoretical formulation of the optical consequences of Lorentz boosts to monochromatic paraxial beams and STWPs. We have shown that monochromatic paraxial beams emitted from a source in a rest frame are converted to a subluminal STWP in a moving frame. Furthermore, Lorentz boost of a subluminal STWP is another subluminal STWP, and the Lorentz boost of a superluminal STWP is another superluminal STWP. Any subluminal STWP can be obtained by a Lorentz boost of a monochromatic beam, and any superluminal STWP can be obtained from an STWP with formally infinite group velocity. We have described an optical system based on spatio-temporal Fourier synthesis that is capable of emulating most of these consequences of Lorentz boost: the change in the group velocity, the temporal bandwidth, the invariant spatial bandwidth, and the concomitant change in the spatio-temporal envelope profile. The only feature not captured by this approach is the Doppler shift of the optical carrier, which does not affect any of the above-listed features. This experimental approach may be of use in emulating Lorentz boosts of optical fields in other contexts, especially in the interaction with moving optical devices.

\section*{Appendix: X-waves}\label{section:transfXwaves}

X-waves are a class of STWPs in which $\omega_{\mathrm{o}}\!\rightarrow\!0$, so that the spatiotemporal spectrum lies at the intersection of the light-cone with a spectral plane of the form:
\begin{equation}\label{eq:XwavePlane}
\frac{\omega}{c}=k_{z}\tan{\theta}.
\end{equation}
This is a propagation-invariant pulsed field with the field given by:
\begin{equation}
E(x,z;t)\!=\!\int\!d\omega\widetilde{E}(\omega)e^{ik_{x}x}e^{-i\omega(t-z/\widetilde{v})}=E(x,0;t-z/\widetilde{v}),
\end{equation}
where the group velocity is $\widetilde{v}\!=\!c\tan{\theta}$, and $\widetilde{E}(\omega)$ is the Fourier transform of $E(0,0;t)$. Note that the entire field is propagation invariant and not only the envelope of the X-wave. A crucial difference between X-waves and baseband STWPs is that the spectral tilt angle $\theta$ is restricted to the range $45^{\circ}\!<\!\theta\!<\!135^{\circ}$, so that X-waves are always superluminal $|\widetilde{v}|\!>\!c$.

Starting with the spectral plane in Eq.~\ref{eq:XwavePlane} for an X-wave in the rest frame $\mathcal{O}$ is transformed into the moving frame $\mathcal{O}'$ under a Lorentz boost to $\tfrac{\omega'}{c}\!=\!k_{z}'\tan{\theta'}$, where $\theta'$ is related to $\theta$ and $\beta$ with the same formula in Eq.~\ref{Eq:TransformingTheta} for STWPs. In other words, the Lorentz boost of an X-wave is another X-wave, and the group velocity is transformed as the velocity of a massive particle. The bandwidth $\Delta\omega$ of the X-wave in $\mathcal{O}$ is changed to $\Delta\omega'$ in $\mathcal{O}'$ via:
\begin{equation}
\frac{\Delta\omega'}{\Delta\omega}=\gamma\left|\frac{\tan\theta-\beta}{\tan\theta}\right|.
\end{equation}



\bibliography{diffraction}
\end{document}